\documentclass[fleqn,usenatbib]{mnras}

\usepackage{newtxtext,newtxmath}

\usepackage[T1]{fontenc}

\DeclareRobustCommand{\VAN}[3]{#2}
\let\VANthebibliography\thebibliography
\def\thebibliography{\DeclareRobustCommand{\VAN}[3]{##3}\VANthebibliography}

\usepackage{graphicx}	
\usepackage{amsmath}	
\usepackage{physics}
\usepackage{url}

\title[$\omega$ Cen Tidal Tails]{Detecting Globular Cluster Tidal Extensions with Bayesian Inference: I. Analysis of $\omega$ Centauri  with Gaia EDR3}

\author[P. B. Kuzma et al.]{
P. B. Kuzma,$^{1}$\thanks{E-mail: pkuzma@roe.ac.uk (PK)}
A. M. N. Ferguson,$^{1}$
J. Pe\~{n}arrubia$^{1}$
\\
$^{1}$Institute for Astronomy, University of Edinburgh, Royal Observatory, Blackford Hill, Edinburgh, EH9 3HJ, UK
}

\date{Accepted XXX. Received YYY; in original form ZZZ}

\pubyear{2021}

\begin{document}
\label{firstpage}
\pagerange{\pageref{firstpage}--\pageref{lastpage}}
\maketitle

\begin{abstract}
The peripheral regions of globular clusters (GCs) are extremely challenging to study due to their low surface brightness nature and the dominance of Milky Way contaminant populations along their sightlines.  We have developed a probabilistic approach to this problem through utilising a mixture model in spatial and proper motion space which separately models the cluster, extra-tidal and contaminant stellar populations.  We demonstrate the efficacy of our method through application to Gaia EDR3 photometry and astrometry in the direction of NGC 5139 ($\omega$ Cen), a highly challenging target on account of its Galactic latitude ($b\approx 15^{\circ}$) and low proper motion contrast with the surrounding field. We recover the spectacular tidal extensions, spanning the $10^{\circ}$ on the sky explored here, seen in earlier work and quantify the star count profile and ellipticity of the system out to a cluster-centric radius of $4^{\circ}$.  We show that both RR Lyrae and blue horizontal branch stars consistent with belonging to $\omega$ Cen are found in the tidal tails, and calculate that these extensions contain at least $\approx 0.1$ per cent of the total stellar mass in the system. Our high probability members provide prime targets for future spectroscopic studies of $\omega$ Cen out to unprecedented radii. 
\end{abstract}

\begin{keywords}
methods: statistical -- stars: kinematics and dynamics -- globular clusters: general -- globular clusters: individual (NGC 5139)-- Galaxy: halo.
\end{keywords}



\section{Introduction}

The outer regions of globular clusters (GCs) are of long-standing interest for a variety of reasons. These regions are shaped by internal processes happening within the star cluster (e.g. two-body relaxation, evaporation), tidal shocking due to disc passages as well as the gravitational potential of the host galaxy in which the cluster orbits.  Moreover, studies of the stellar kinematics in the peripheral regions of GCs have the potential to shed light on whether they are embedded in dark matter mini-halos, which is of critical relevance to understanding their origins  \citep[e.g.,][]{1984ApJ...277..470P, 2017MNRAS.471L..31P}. On the other hand, knowledge of the chemical properties of stars in the far outer regions of GCs is important for understanding what drives the multiple population phenomenon \citep[e.g.,][and references therein]{2018ARA&A..56...83B} and for constraining how much of the Galactic halo can be composed of their tidally-stripped stars  \citep[e.g.,][]{2010A&A...519A..14M, 2019A&A...625A..75K}. 

Prior to 2018, our understanding of GC outskirts had been slowly pieced together from an assortment of studies and surveys \citep[e.g.,][]{ 2003AJ....126.2385O, 2006ApJ...637L..29B, 2009AJ....138.1570O, 2014MNRAS.445.2971C, 2016MNRAS.461.3639K}. This work  was extremely challenging due to the very low stellar densities in these parts
and the fact that usually only photometry was available to disentangle genuine GC stars from the dominant population of field contaminants.  Nonetheless, these pioneering studies demonstrated that extended structures were not unusual around GCs. In some cases, these structures take the form of narrow extended tails which can span many tens of degrees on the sky, features which are entirely expected for a star cluster orbiting within a tidal field \citep[e.g.,][]{2010MNRAS.401..105K}. In other cases, diffuse stellar envelopes are seen which can be traced to at least a few hundred parsecs in radius around the cluster \citep[e.g.,][]{2018MNRAS.473.2881K}. These envelopes  may also be the result of dynamical evolution within a tidal field, consisting largely of potential escaper stars \citep{2001ASPC..228...29H, 2017MNRAS.468.1453D, 2019MNRAS.487..147C}, but another tantalizing possibility is that they are the remnants of accreted dwarf galaxies in which the GCs were once the nuclei \citep[e.g.,][]{1988IAUS..126..603Z}. In this scenario, diffuse stellar envelopes would signify a definite extragalactic origin for a GC and establishing the numbers of clusters with these types of structures would provide key information for piecing together the assembly history of the Galaxy.  

Like most other aspects of Galactic astronomy, the study of GCs has been revolutionised with the second data release from the Gaia mission \citep[DR2;][]{2018A&A...616A...1G} and, most recently, the third early data release \cite[EDR3;][]{2016A&A...595A...1G,2020arXiv201201533G}. DR2 and EDR3 provide precision photometry and, of particular importance, astrometry for over a billion Milky Way (MW) stars. The availability of parallaxes and proper motions enable an enormous improvement in the ability to isolate even very tenuous groups of co-moving stars from the general MW field. Indeed, both data releases have been used to search for new halo substructures in the MW with great success. For example, DR2 has uncovered evidence for significant merger events in the early history of the Galaxy \citep{2018Natur.563...85H, 2019MNRAS.488.1235M}, and a wealth of new stellar streams in the halo  \citep[e.g.,][]{2019ApJ...872..152I}.   Furthermore, our understanding of previously known stellar streams,  such as the GD-1 and the Orphan streams, has greatly improved thanks to the DR2 astrometry  \citep[e.g.,][]{2018ApJ...863L..20P,2019MNRAS.485.4726K,2019MNRAS.486..936F}. 

Gaia DR2 has enabled the proper motions and orbits for the entire GC population to be calculated \citep[e.g.,][]{2018A&A...616A..12G,2019MNRAS.484.2832V,2020arXiv200813624B}, and their structural parameters have also been explored \citep{2019MNRAS.485.4906D}. Even the kinematics of the inner regions of GCs have been investigated, such as a radial velocity dispersion profiles and rotation \citep{2019MNRAS.485.1460S,2019MNRAS.482.5138B,2020A&A...638L..12P}. While the majority of studies have focused on the bright main bodies of GCs, DR2 has also led to dramatic improvements in our ability to study GC peripheries.  Since the release of DR2, faint extra-tidal features have been discovered surrounding a number of clusters, such as E3 \citep{2020MNRAS.499.2157C}, NGC 362 \citep{2019MNRAS.486.1667C}, NGC 7099 \citep{2020MNRAS.495.2222S,2020A&A...643A..15P}, while a number of other clusters have had tidal extensions either recovered or found to extend further than previously known \cite[e.g,][]{2020MNRAS.495.2222S,2020ApJ...889...70B}. 

One cluster that has particularly benefited from this new wealth of information is NGC 5139 ($\omega$ Centauri, hereafter $\omega$ Cen), the most massive GC in the MW and certainly one of the most peculiar in terms of its properties. Studies over the last few decades have revealed complex stellar populations in this system which not only display a spread in many light elements \citep[e.g.,][]{2010ApJ...722.1373J,2017MNRAS.469..800M} but also in age and [Fe/H]  \citep{1995ApJ...447..680N,2014ApJ...791..107V}. The fact that $\omega$ Cen is also on a tightly-bound retrograde orbit about the MW \citep{1999AJ....117..277D} has led to suggestions that it did not form in-situ and a long-held belief that $\omega$ Cen it may actually be the core of a now defunct dwarf galaxy accreted in the early history of the MW \citep{2000LIACo..35..619M, 2003MNRAS.346L..11B}.   A number of pre-Gaia studies endeavoured to search for tidally-stripped material around $\omega$ Cen, the properties of which could be able to confirm this scenario.   Using star counts measured on deep photographic plates,  \citet{2000A&A...359..907L} found evidence for the presence of tidal extensions extending north and south of the cluster however this structure was later shown to be due to variable extinction \citep{2003AJ....126.1871L}. Further photometric as well as spectroscopic searches yielded null or only marginal results \citep[e.g.,][]{2008AJ....136..506D, 2009MNRAS.396.2183S,2015A&A...574A..15F}.

Studies of the peripheral regions of $\omega$ Cen are greatly hindered by its proximity to the Galactic Plane ($b\approx 15^{\circ}$),  resulting in a very significant foreground/background contamination along its sightline, as well as significant variable extinction.  The power of Gaia DR2 to provide a first glimpse of its outer regions was demonstrated in a spectacular fashion by \cite{2019NatAs.tmp..258I}. Motivated by the similarity in orbital properties of the \textit{Fimbulthul} stellar stream and $\omega$ Cen \citep{2019ApJ...872..152I}, these authors used N-body models to conduct a tailored search for debris in the vicinity of the cluster and discovered tidal extensions extending several degrees across the sky. Using a more detailed mixture modelling approach,  \cite{2020MNRAS.495.2222S} has also 
recovered these extensions.  

In spite of these promising developments, the origin and evolutionary history of  $\omega$ Cen remain uncertain. For example, on the basis of orbital properties, \cite{2019MNRAS.488.1235M} argue that  $\omega$ Cen is a GC originally accreted along with the Sequoia galaxy while  \citet{2019A&A...630L...4M} and \citet{2021ApJ...909L..26B} instead argue for a link to the \textit{Gaia}-Enceladus accretion event.  Folding in age and metallicity information also fails to settle this debate \citep{2020MNRAS.498.2472K, 2020MNRAS.493..847F}.  Furthermore, while the evidence that $\omega$ Cen is the progenitor of the  \textit{Fimbulthul} stream is compelling in terms of their orbits \citep{2019NatAs.tmp..258I,2019ApJ...872..152I}, the chemical link rests on only two stars  \citep{2020MNRAS.491.3374S}. 

In this paper, we demonstrate the efficacy of a new probabilistic technique that we have developed to explore GC peripheries through application to Gaia EDR3 data in the vicinity of $\omega$ Cen. Our mixture model approach invokes physically-motivated models for the proper motions and the spatial distributions of the cluster, extra-tidal  and contaminant populations, and is solved within a Bayesian framework. 
In Sec. \ref{sec:data} we discuss the dataset and in Sec. \ref{sec:method} we present our methodology.  In Sec. \ref{sec:resdisc} we present our successful recovery of the tidal tails seen in earlier work and conduct a detailed analysis of their properties, including comparison to other stellar population tracers and existing radial velocity data. 

\section{The Data} \label{sec:data}
\subsection{Initial Selection}
We base our analysis on high-quality proper motions and photometry provided by the Gaia mission, and in particular EDR3 \citep{2016A&A...595A...1G,2020arXiv201201533G}. We began by retrieving all stars within a five degree radius of $\omega$ Cen from the Gaia archive\footnote{https://gea.esac.esa.int/archive/} using Table Access Protocol (TAP) facilities. With our adopted distance to the cluster of 5.2 kpc \citep{1996yCat.7195....0H,2021ApJ...908L...5S}, this corresponds to a physical radius of $\sim$450 pc. We first applied the several adjustments to the data as suggested by the EDR3  documentation. Specifically, these are the zero-point correction in the parallax using the code provided in \cite{2020arXiv201201742L} and the correction of the $G$-band magnitude and the corrected flux excess factor as provided in \cite{2020arXiv201201916R}.  We then applied a series of cuts to the resulting catalogue, namely:

\begin{itemize}
\item{stars with $(G_{BP}-G_{RP})> 1.6$~mag are excised as this part of the colour magnitude diagram (CMD) primarily contains field dwarfs;}
\item{due to overestimation of the $G_{BP}$ flux causing sources to appear to blue, we removed all stars $G_{BP}>20.3$ mag \citep[see Section 8.1 in][]{2020arXiv201201916R}};
\item{stars whose corrected BP and RP excess flux is greater than 3 times the associated uncertainty  \citep[see Section 9.4 in][]{2020arXiv201201916R}} are removed as this typically relates to poor photometry;
\item{stars with poor astrometric data are removed, using the `Re-normalised Unit Weight Error'
, $u_r > 1.4$.  This value has been shown to provide the optimal selection of stars with `good' astrometric solutions \citep{2018A&A...616A...2L, 2020arXiv201201742L}}
\item{stars are removed that have a well-measured parallax that places them within 3 kpc of the Sun: $ (\varpi - 3\sigma_{\varpi} )> 0.3$ mas. This removes all nearby stars that contaminate the $\omega$ Cen sightline.}
\end{itemize}

These selections pruned the original sample of 2.5 million stars to 1.7 million stars, a removal of 30 per cent. We then transformed all positions and proper motions onto a tangential coordinate projection about the centre of $\omega$ Cen using the equations (2) in \cite{2018A&A...616A..12G} - ($\xi$,$\eta$) now denote positions (for $\alpha$ and $\delta$, respectively) and $\mu^{*}_{\xi}$ and $\mu_{\eta}$ denote the associated proper motions. In addition, we also corrected the proper motion of all stars in the sample for the solar reflex motion at the adopted distance of 5.2~kpc using the python \textsc{gala} package \citep{gala}, which uses the solar motion $(12.9, 245.6, 7.78)$ km s$^{-1}$ \citep{2018RNAAS...2..210D}. Any stars that do not have a proper motion measurement in EDR3 are also removed at this stage.

For the subsequent analysis, we estimate EDR3 photometric uncertainties from the flux zero-point equations given in \cite{2020arXiv201201916R}.  Furthermore, the photometry is de-reddened using the \citet{2011ApJ...737..103S} dust maps and the colour-dependent extinction equations given in \citet{2018A&A...616A..10G}. Given the large area of the sky spanned by our dataset and the highly variable extinction across this region (see Fig. \ref{fig:dust}), the de-reddening correction has been performed on a star-by-star basis, and the resultant de-reddened photometry is denoted as $G_0$, $G_{BP,0}$ and $G_{RP,0}$. 
 
 \begin{figure}
\setcounter{figure}{0}
   \begin{center}  
    \includegraphics[width=\columnwidth]{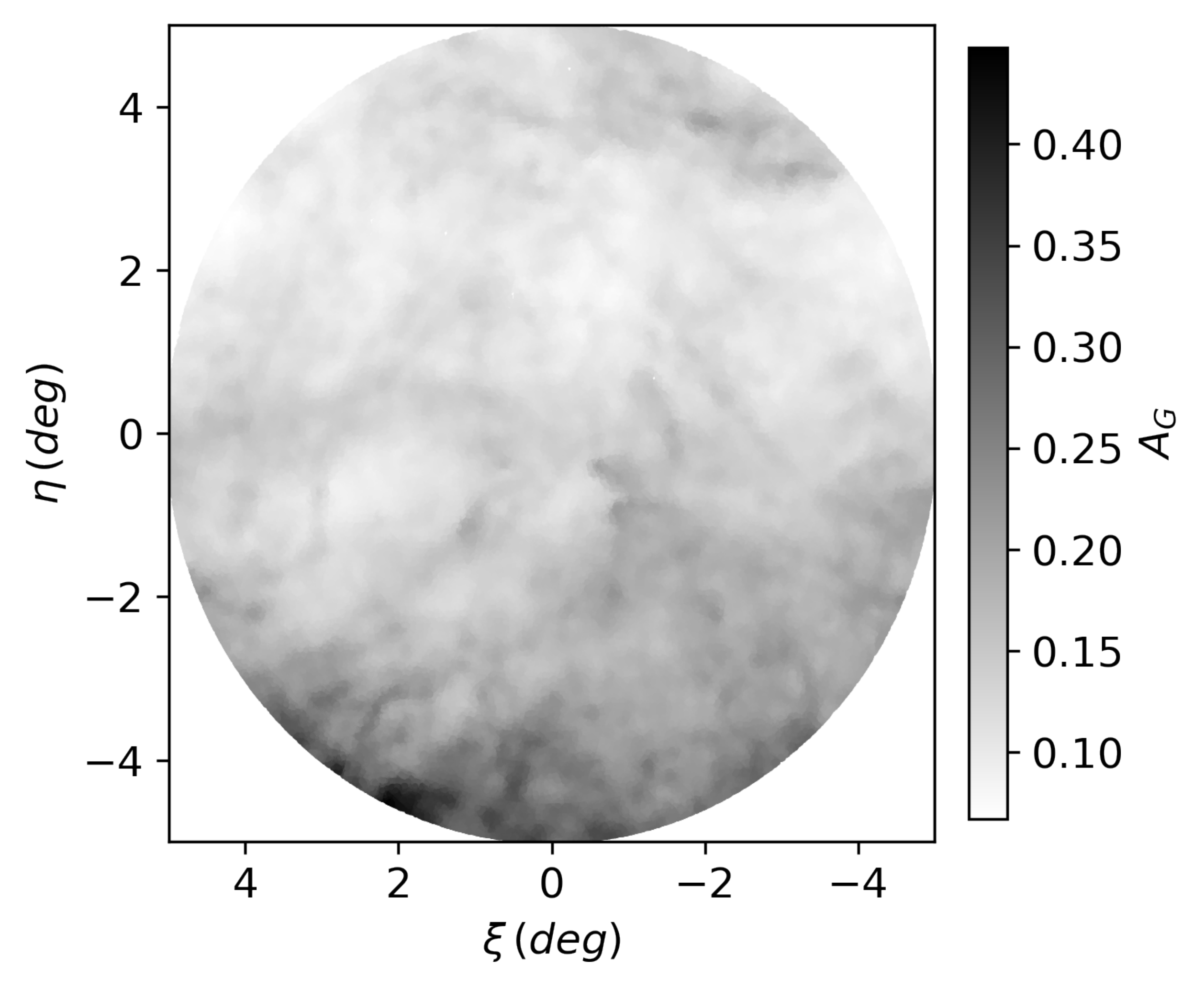}
  \end{center}
\caption{Extinction map across our retrieved field of view from \citet{2011ApJ...737..103S}. The extinction is variable across the field of view, but most severe in the south-east and south directions, in the direction of the Galactic Plane.}
 \label{fig:dust}
\end{figure}

 \subsection{Photometric Selection}
\label{sec:photsel}
\begin{figure}
  \begin{center}  
    \includegraphics[width=\columnwidth]{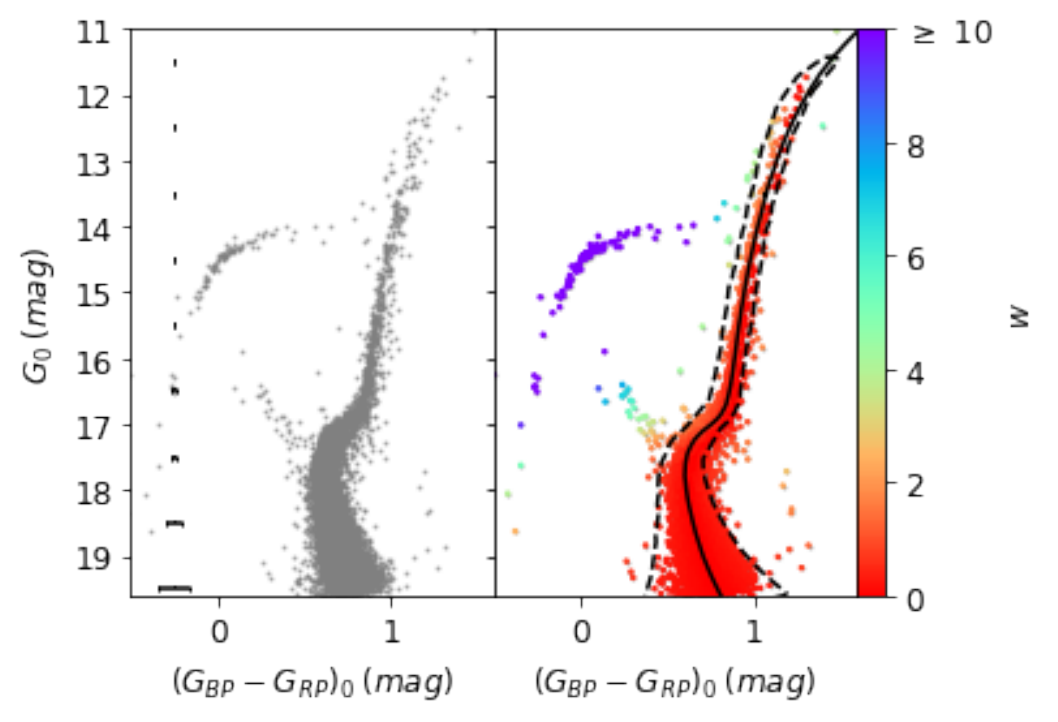}
  \end{center}
\caption{CMD of the fiducial sample stars of $\omega$ Cen. Left: CMD with associated photometric uncertainties after initial selection. Right: Same plot showing the addition of the Dartmouth isochrone ([Fe/H]$=-1.53$ dex, [$\alpha$/Fe]$=+0.2$) and the results of applying our photometric CMD selection, with stars coloured based on their $w$ value. The dashed-line polygon indicates the boundary of the photometric selection.}
\label{fig:CMD}
\end{figure}

After the initial cuts on the data were performed, we proceeded to remove further contaminants from our sample through additional photometric selection. Specifically, we wanted to retain only those stars that were consistent with the well-defined sequence for $\omega$ Cen in colour-magnitude space (Fig \ref{fig:CMD}, left). Because incompleteness due to crowding affects the quality and depth of the photometry in central regions of the cluster, we defined a  `fiducial sample' of stars that lie within a cluster-centric radius of 10 to 50 arcmin. The outer radius of this sample is just beyond the nominal King tidal radius of 70.25~pc presented in \cite{2019MNRAS.485.4906D}, which corresponds to 46.4 arcmin at our adopted distance.
The proper motion of $\omega$ Cen has been derived to be ($-3.25, -6.75$) mas yr$^{-1}$ by \cite{2021mnras210209568V} using Gaia EDR3. We refine the fiducial sample by removing stars with proper motions that differ by more than 1 mas yr$^{-1}$ of this value when transformed to the solar reflex motion-corrected frame:  $(\mu^*_{\alpha},\mu_{\delta})$=($3.08, -3.58$) mas yr$^{-1}$. To guide our photometric selection, we fit a PARSEC\footnote{\url{http://stev.oapd.inaf.it/cmd}} isochrone \citep{2012MNRAS.427..127B,2017ApJ...835...77M} with [Fe/H]$=-1.53$ dex, [$\alpha$/Fe]$=+0.2$ and an age of 12 Gyr in the Gaia EDR3 bandpasses to the fidicual sample. Although it is established that multiple stellar populations exist within $\omega$ Cen which results in significant colour width on the RGB,  these are believed to be largely concentrated 
within the central 10 arcmin \citep[e.g.,][]{2009A&A...507.1393B} and we assume that they have negligible effect in the outermost regions that we focus on here. 
We assign each star in the fiducial sample a pseudo-colour value, $w$, which is defined as the absolute difference between the  $(G_{BP}-G_{RP})_0$ colour of the star and that of the isochrone at the corresponding $G_0$, which we denote as $(G_{BP}-G_{RP})_{ISO}$, divided by the star's uncertainty in colour $\sigma_{(G_{BP}-G_{RP})0}$ (see also \citealt{2018MNRAS.473.2881K}). That is:

\begin{equation}
w=\abs{\frac{(G_{BP}-G_{RP})_0-(G_{BP}-G_{RP})_{ISO}}{\sigma_{(G_{BP}-G_{RP})_0}}}.
\end{equation}
\noindent
Stars that are photometrically-identical to the mean fiducial of $\omega$ Cen thus have $w=0$ and this value increases when the colour difference is large relative to the photometric uncertainty.  

We determine the distribution of $w$ as function of $G_0$ to identify the best $w$ range that describes the cluster sequence, assuming a half-normal distribution. At a given $G_0$, we record the standard deviation of the distribution ($\sigma_{w}$) and adopt $2\sigma_{w}$ to capture the bulk of the stellar sequence. The function $\sigma_2(G_0)$ then represents how $2\sigma_{w}$ varies as a function of $G_{0}$. Due to systematic effects,  $\sigma_2(G_0)$ increases significantly at the bright end (i.e. as $G_{0}$ and the photometric uncertainties decrease) so we employ an exponential fit to $\sigma_2(G_0)$ to allow for this effect. This keeps the red giant branch (RGB) sufficiently covered while selecting almost all stars on the main sequence and the turn-off regions, minimizing the contamination from MW field stars. Finally, with $\sigma_2(G_0)$ defined, we calculate $w$ for stars within the entire (5$^{\circ}$ radius) sample, and keep only stars that satisfy the condition $w \leq \sigma_2(G_0)$. This removes a further 1.2 million stars from the sample, and yields a final sample of $5\times10^{5}$ stars that we feed to our model. The fitted isochrone. the $w$ range and the boundary of the $\sigma_2(G_0)$ cut are shown in Fig. \ref{fig:CMD} (right). We note that this selection excludes blue horizontal branch (BHB) stars and RR-Lyrae (RRL) stars at this point in the analysis but we will revisit those populations in a later section. 

\section{The Method}\label{sec:method}

Our goal is to identify stars in the very outer regions of $\omega$ Cen, well beyond the nominal tidal radius, that are potentially members of the cluster. 
Given the dominant and spatially-varying contaminant population of MW foreground and background stars along the line-of-sight to the cluster, this is best done using a probabilistic approach since hard cuts on astrometric parameters may struggle to cleanly isolate the expected weak signal. Another advantage of such an approach is that it can also yield high-priority candidates for future spectroscopic follow-up, which will be an obvious next step in exploring GC peripheries.  

We adopt a maximum-likelihood procedure,  first presented in \cite{2020IAUS..351..468K}, that is inspired by the earlier work of \citet{2009AJ....137.3109W} and \citet{2011ApJ...742...20W} who use it to derive stellar membership probabilities for MW dwarf spheroidal galaxies. Similar work has  been pursued more recently by \citet{2019ApJ...875...77P} and \citet{2020AJ....160..124M}, who apply the methodology to derive the proper motions of ultra-faint dwarf systems with Gaia DR2 data. In essence, we model the spatial distribution and proper motions of stars with a Gaussian mixture model and solve for the parameters within a Bayesian framework. 

Our mixture model consists of three components -- one representing the cluster proper ($cl$), one representing the cluster extra-tidal structure ($ex$) and the other the MW contaminant population ($MW$). The total likelihood $\mathcal{L}_{tot}$, function can be expressed as: 
\begin{equation}
\mathcal{L}_{tot}=f_{cl+ex}(f_{cl}\mathcal{L}_{cl}+(1-f_{cl})\mathcal{L}_{ex})+(1-f_{cl+ex})\mathcal{L}_{MW}\label{eq:ltot}
\end{equation}
\noindent
where $\mathcal{L}_{cl/MW/ex}$ correspond to the likelihoods of the cluster, contaminant and extended structure components respectively and $f_{cl/cl+ex}$ correspond to the fraction of stars in the $cl$ and $cl+ex$ components. These latter parameters are defined as:

\begin{equation}
f_{cl+ex}=\frac{N_{cl}+N_{ex}}{N_{cl}+N_{ex}+N_{MW}}
\end{equation}
\noindent
and
\begin{equation}
f_{cl}=\frac{N_{cl}}{N_{cl}+N_{ex}}
\noindent
\end{equation}
\noindent 
where $N$ is the total number of stars in that component. 

Each component in the total likelihood function, $\mathcal{L}_{cl/MW/ex}$ takes the following form:
\begin{equation}
\mathcal{L}_{cl/MW/ex}=\mathcal{L}_{pm}\mathcal{L}_{spat}\label{eq:lcom}
\end{equation}
\noindent
where $\mathcal{L}_{pm}$ and $\mathcal{L}_{spat}$ are the likelihood functions for the proper motion and the spatial distribution, respectively.  For each component of our mixture model, the proper motion likelihood takes the form of a multivariate Gaussian:
\begin{equation}
\ln{\mathcal{L}_{pm}}=-\frac{1}{2}(X-\bar{X})^\top\,C^{-1}\,(X-\bar{X}) - \frac{1}{2}\ln{4\pi\det[C]} \label{eq:lpm}
\end{equation}
\noindent
where $\bar{X} = (\bar{\mu^{*}_{\xi}},\bar{\mu_{\eta}})$ is the systemic proper motion of the given component, and $X=({\mu^{*}_{\xi}},{\mu_{\eta}})$ is the data vector. The covariance matrix, $C$, is given by:
\begin{equation}
C=\begin{bmatrix}
\sigma^2_{\mu^{*}_{\xi}}+\sigma^2_{\xi,cl/MW}& \rho\sigma_{\mu^{*}_{\xi}} \sigma_{\mu_{\eta}}\\
\rho\sigma_{\mu^{*}_{\xi}} \sigma_{\mu_{\eta}}& \sigma^2_{\mu_{\eta}}+\sigma^2_{\eta,cl/MW}\\
\end{bmatrix}. \label{eq:cov}
\end{equation}
\noindent
where $(\sigma_{\mu^{*}_{\xi}},  \sigma_{\mu_{\eta}})$ denote the proper motion uncertainties from Gaia,  $(\sigma_{\xi,cl/MW}, \sigma_{\eta,cl/MW})$  denote the proper motion dispersion of the cluster and MW components  and $\rho$ denotes the correlation between $\mu^{*}_{\xi}$ and $\mu_{\eta}$. As we are searching for outlying stars that belong to the cluster, the proper motion for the cluster and extended structure components are assumed to be identical. This is a reasonable assumption since the proper motion should not vary significantly over the area on the sky that we are exploring \citep[e.g.,][]{2016ApJ...833...31B, 2019ApJ...887L..12B}. We also fit the dispersion of the MW contaminant and cluster components. In the latter case, a dispersion is included to compensate for crowding issues that increase the uncertainty in the proper motions in inner regions of the cluster; this dispersion is set to zero in the extended structure as crowding is no longer an issue.  In total, the proper motion components of our model contribute eight free parameters: the proper motion and proper motion dispersion in both the $\xi$ and $\eta$ directions for each of the field and the GC+extended structure components. 

The spatial distribution component of the likelihood function, $\mathcal{L}_{spat}$, is expressed in terms of the tangential projection of stellar positions about the cluster centre.  As will become apparent, our analysis is simplified if we use polar coordinates ($r,\theta$) and, following \citet{2011ApJ...742...20W}, we define this as:

\begin{equation}
\mathcal{L}_{spat}(R,\Theta)=\pdv{}{R}{\Theta}\frac{\int_{0}^{R}\int_{0}^{\Theta} r \Sigma(r,\theta) \partial r \partial\theta}{\int_{0}^{R_{max}}\int_{0}^{2\pi} r \Sigma(r,\theta) \partial r \partial\theta}
\end{equation}
\noindent
where $\Sigma(r,\theta)$ is the stellar surface density of the component in question, $R_{max}$ is the radius of the field of view, and $(R,\Theta)$ is the data vector containing the cluster-centric radius and position angle for each star.  For the cluster component,  we adopt a \cite{1962AJ.....67..471K} model for simplicity:
\begin{equation}
 \Sigma_{cl}(r)=\Sigma_{0}\left(\frac{1}{\sqrt[]{1+r^2/r^2_{c}}}-\frac{1}{\sqrt[]{1+r^2_{t}/r^2_{c}}}\right)^2
 \end{equation}
\noindent
where $\Sigma_{0}$ is the central surface brightness, $r_c$ and $r_t$ are the King core and tidal radii. Here, we sample a Gaussian distribution using the measurements and uncertainties for $r_c$ ($2.34\pm0.09$ arcmin) and $r_t$ from \cite{2005ApJS..161..304M} and \cite{2019MNRAS.485.4906D}, respectively. As we are ultimately interested in searching the periphery of $\omega$ Cen for potential members, the structural parameters of the central regions are not considered to be of primary importance and hence are not fit in our analysis.

The large angular region we are exploring around $\omega$ Cen, combined with its low Galactic latitude, means that we cannot assume that the spatial distribution of the MW contaminant population is constant across the field. Instead, we assume a linear model for this component, given by:
\begin{equation}
 \Sigma_{MW}(r,\theta)=1+k_{MW}\cos{(\theta-\theta_{MW})}\label{eq:linspat}
 \end{equation}
 \noindent
 where $k_{MW}$ is the magnitude of the gradient and  $\theta_{MW}$ is its direction. 

Finally,  in modelling the extended component, we want to be sensitive to both axisymmetric extensions, such as tidal tails, as well as 
circularly-symmetric envelopes. This flexibility can be achieved by adopting a quadrupole model for the extended component: 
\begin{equation}
\Sigma_{ex}(r,\theta)=r^{-\gamma}(1+k_{ex}\cos^2{(\theta-\theta_{ex})})\label{eq:qudspat} 
\end{equation}
\noindent
where $k_{ex}$ is the magnitude of the quadrupole, $\theta_{ex}$ is its direction and $\gamma$ is the power-law index. The free parameters in this equation can be used to classify whether a feature is detected, and whether that feature is in the form of axisymmetric tidal tails, a diffuse stellar envelope or somewhere in between. For example, if $\theta_{ex}$ is returned as uncertain, this implies the lack of a preferred orientation to the extended structure. Along with a well-determined $\gamma$, this would indicate the detection of a spherical envelope.  
Overall, the spatial components contribute five free parameters: the magnitude and position angle of the extended structure and MW contaminant components, and the power-law index of the extended structure. 

The five free parameters from the spatial components, combined with the eight free parameters from the proper motion components, and the two normalization factors 
$(f_{cl+ex}, f_{cl})$, mean that we have a total of 15 free parameters. In fitting these, we assume the following priors:

\begin{itemize}
\item{Linear priors between -10 and 10 ${\rm mas \,yr}^{-1}$ for the proper motion components $\mu^{*}_{\xi}$ and $\mu_{\eta}$ of both the cluster+extended structure and the MW contamination;}
\item{Log priors between $-3$ and $2$ ${\rm mas\,yr}^{-1}$ for the proper motion dispersion $\sigma_{\mu^{*}_{\xi}}$ and $\sigma_{\mu_{\eta}}$ of both the cluster and the MW contamination;}
\item{Log priors between $-4$ and $5$ for the linear gradient of the contaminant distribution, $k_{MW}$, and for the quadrupole strength of the extended component,  $k_{ex}$. Linear priors between $0^{\circ}$ and $360^{\circ}$ for the position angle of the MW contaminant distribution, $\theta_{MW}$, and between $0^{\circ}$ and $180^{\circ}$ for the position angle of the extended structure component $\theta_{ex}$.}
\item{Linear priors between 0 and 1 for the two normalization factors, $f_{cl+ex}$ and $f_{cl}$.}
\item{A linear prior between 0 and 2 for the power law index of the extended component, $\gamma$.}
\end{itemize}

The Python wrapper for the Bayesian inference tool MultiNest, PyMultiNest, is used to determine the posterior distributions \citep{2014A&A...564A.125B}. By sampling the entire posterior distribution, we can assign to each star in our sample a probability that it is a member of the cluster + extended structure component, $P_{mem}$. This is computed as the ratio of the GC + extended structure likelihoods to the total likelihood, or:
\begin{equation}
P_{mem}=\frac{f_{cl+ex}(f_{cl}\mathcal{L}_{cl}+(1-f_{cl})\mathcal{L}_{ex})}{\mathcal{L}_{tot}}.
\end{equation}
\noindent
We define the mean $P_{mem}$ from the sampling as our membership probability for each star. We designate all stars with $P_{mem}\,\geq 0.5$ as highly-probable members,  Stars with $P_{mem}<0.5$ are low probability members or contaminants and are not considered further in this paper.

\begin{table*}
\begin{center}
\caption{Summary of the results of our Bayesian nested sampling and a brief description of the parameters. The proper motions presented here have been corrected for solar reflex motion.}
\label{tab:SDSS}
\begin{tabular}{@{}ccccc}
\hline \hline
Parameter&\multicolumn{2}{c}{Prior}&Posterior (Units)&Description\\
&Min&Max&1$\sigma$ confidence&\\

\hline

$\mu^{*}_{\xi,cl}$&-10&10& $3.055 \pm 0.002$ (mas yr$^{-1}$)& Proper motion of the cluster in the $\xi$ direction.\\
$\mu_{\eta,cl}$&-10&10&$-3.624\pm 0.002$ (mas yr$^{-1}$)&Proper motion of the cluster in the $\eta$ direction.\\
$\sigma_{\mu^{*}_{\xi,cl}}$&$10^{-3}$&$10^{2}$&$0.394 \pm 0.002$ (mas yr$^{-1}$)&  Proper motion dispersion of the cluster in the $\xi$ direction. \\
$\sigma_{\mu_{\eta},cl}$&$10^{-3}$&$10^{2}$&$0.378 \pm 0.002$  (mas yr$^{-1}$)&  Proper motion dispersion of the cluster in the $\eta$ direction. \\
$\mu^{*}_{\xi,MW}$&-10&10&$0.401 \pm 0.005$ (mas yr$^{-1}$)&  Proper motion of the MW field in the $\xi$ direction. \\
$\mu_{\eta,MW}$&-10&10&$1.579 \pm 0.003$ (mas yr$^{-1}$)&  Proper motion of the MW field in the $\eta$ direction. \\
$\sigma_{\mu^{*}_{\xi,MW}}$&$10^{-3}$&$10^{2}$&$3.206 \pm 0.004$ (mas yr$^{-1}$)&  Proper motion dispersion of the MW field in the $\eta$ direction. \\
$\sigma_{\mu_{\eta},MW}$&$10^{-3}$&$10^{2}$&$1.992 \pm 0.002$ (mas yr$^{-1}$)&  Proper motion dispersion of the MW field in the $\xi$ direction. \\
$f_{cl+ex}$&0&1&$0.104 \pm 0.0004$&  Normalization constant between the cluster+extended structure and the field .\\
$f_{cl}$&0&1&$0.910 \pm 0.002$ &  Normalization constant between the cluster and extended structure. \\
$\theta_{MW}$&$0$&$360$&$147.852^{\circ} \pm 0.347^{\circ}$  &  Position angle of the linear gradient describing the field. \\
$k_{MW}$&$10^{-4}$&$10^{5}$&$ 0.349 \pm 0.002$&  Magnitude of the linear gradient describing the field. \\
$\theta_{ex}$&$0$&$180$&$122.878^{\circ} \pm 2.141^{\circ}$&  Position angle of the quadrupole describing the extended structure. \\
$k_{ex}$&$10^{-4}$&$10^{5}$&$1.602 \pm 0.201$&  Magnitude of the quadrupole describing the extended structure. \\
$\gamma$&0&2&$1.549 \pm 0.007$&   Power-law index of the radial profile describing the extended structure.\\

\hline
\end{tabular}
\end{center}

\end{table*}

\section{Results and Discussion}\label{sec:resdisc}
\subsection{The tidal extensions of \texorpdfstring{$\omega\,$}{}Cen}\label{sect:tidal_extensions}
Our Bayesian technique has measured the posterior distributions for 15 different parameters, and the solutions are presented in Table \ref{tab:SDSS} (the posterior distributions are presented as on-line material).  The non-zero normalisation constants derived for $f_{cl+ex}$ and $f_{cl}$ indicate that our analysis has recovered the cluster as well as detected an extended extra-tidal feature surrounding it. 

\begin{figure}
  \begin{center}  
    \includegraphics[width=\columnwidth]{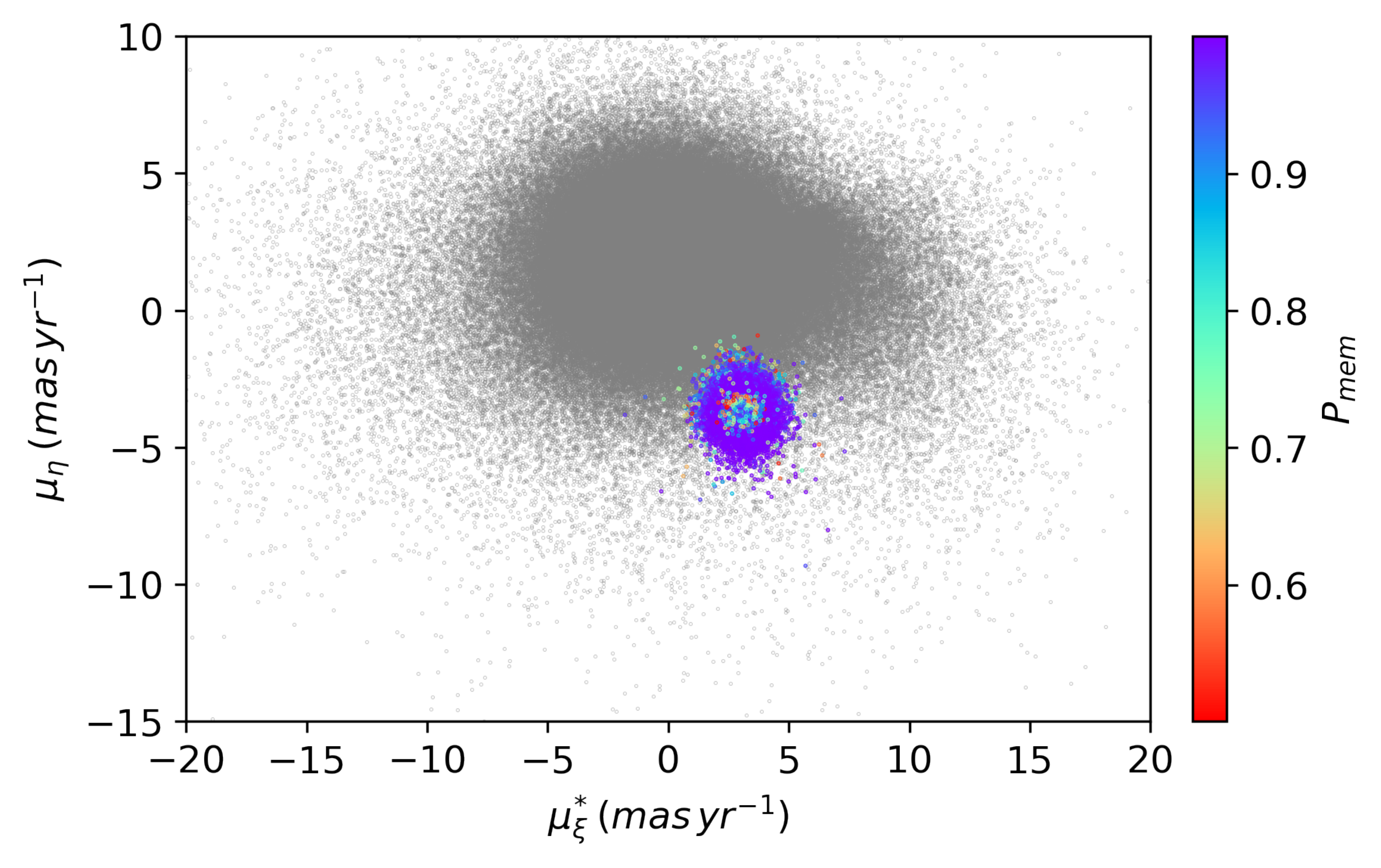}
  \end{center}
\caption{Tangentially-projected proper motion distribution corrected for solar reflex motion. Stars with $P_{mem} \geq0.5$ are coloured according to their membership probability, while those with $P_{mem} <0.3$ are shown in grey. The proper motion signature of $\omega$ Cen is clearly visible and well-defined by the high probability members. }
\label{fig:PM}
\end{figure}

\begin{figure}
  \begin{center}  
    \includegraphics[width=\columnwidth]{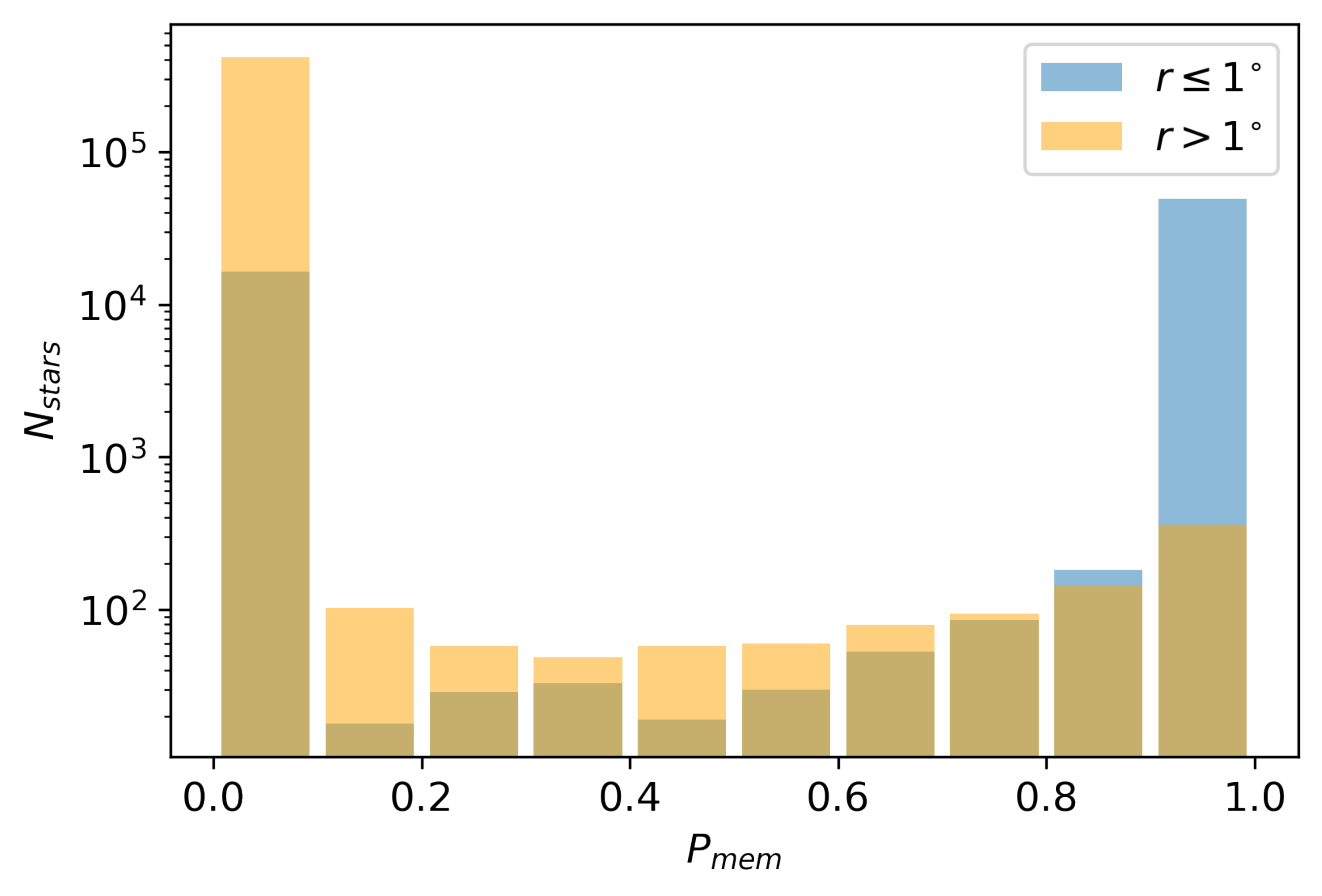}
  \end{center}
\caption{Histogram of membership probability, $P_{mem}$, for stars within a 1 deg radius in blue, and outside of this radius in yellow. This demonstrates the effectiveness of  our technique for distinguishing between cluster and foreground stars at both small and large radii. }
\label{fig:pmeme_hist}
\end{figure}

In Fig. \ref{fig:PM}, we present the proper motion distribution of our sample, highlighting those stars with an $\omega$ Cen membership probability of $\geq0.5$.  The cluster clearly stands out relative to the dominant MW field distribution. We fit the solar reflex motion corrected proper motion  of $\omega$ Cen to be (${\mu^{*}_{\xi}},\mu_{\eta})=(3.055\pm0.002,-3.624\pm 0.002)$ mas yr$^{-1}$. This corresponds to (${\mu^{*}_{\xi}},\mu_{\eta})=(-3.21, -6.74)$ in the observed frame and is in excellent agreement with the recently determined EDR3 values from \citet{2021mnras210209568V}. We find that $f_{cl+ex}\approx 0.10$, which indicates that ${\sim}10$ per cent of our photometrically-selected stars belong to the cluster + extended structure components on the basis of their proper motions and spatial positions, while the remaining ${\sim}90$ per cent belong to the MW field. This high level of MW contamination is not surprising given the low Galactic latitude of  $\omega$ Cen and the large area on the sky that we have analysed.  Nonetheless, the membership probability distribution presented in Fig. \ref{fig:pmeme_hist} demonstrates that our method is effective in distinguishing between cluster and contaminant stars at both small and large cluster-centric radii, with a couple thousand high probability members lying beyond the King tidal radius.

We unambiguously detect an extra-tidal component to  $\omega$ Cen in the form of compelling tidal tails which span the $\sim 10^{\circ}$ on the sky that we have analysed. The normalization factor between the cluster and the extended structure is $f_{cl}=0.910 \pm 0.002$, indicating that ${\sim}91$ per cent of the stars we assign to the cluster reside within the King tidal radius, while ${\sim}9$ per cent lie in the tidal extension. However, we stress that this is not representative of the overall fractional mass or luminosity in the extended structure component since photometric completeness issues at small and large radii, as well as the stellar mass function of stars in the different components, are complicating issues (see Section \ref{sect:radial_profile}). The position angle of the extended component is well defined,  $\theta_{ex}=122.88^{\circ} \pm 2.14^{\circ}$ (East of North), indicating a preferred orientation. Notably, this is not in the same direction as the direction of the MW field gradient, which is $\theta_{MW}=147.85^{\circ} \pm 0.35^{\circ}$ (East of North). 
\begin{figure*}
  \begin{center}  
    \includegraphics[width=0.95\textwidth]{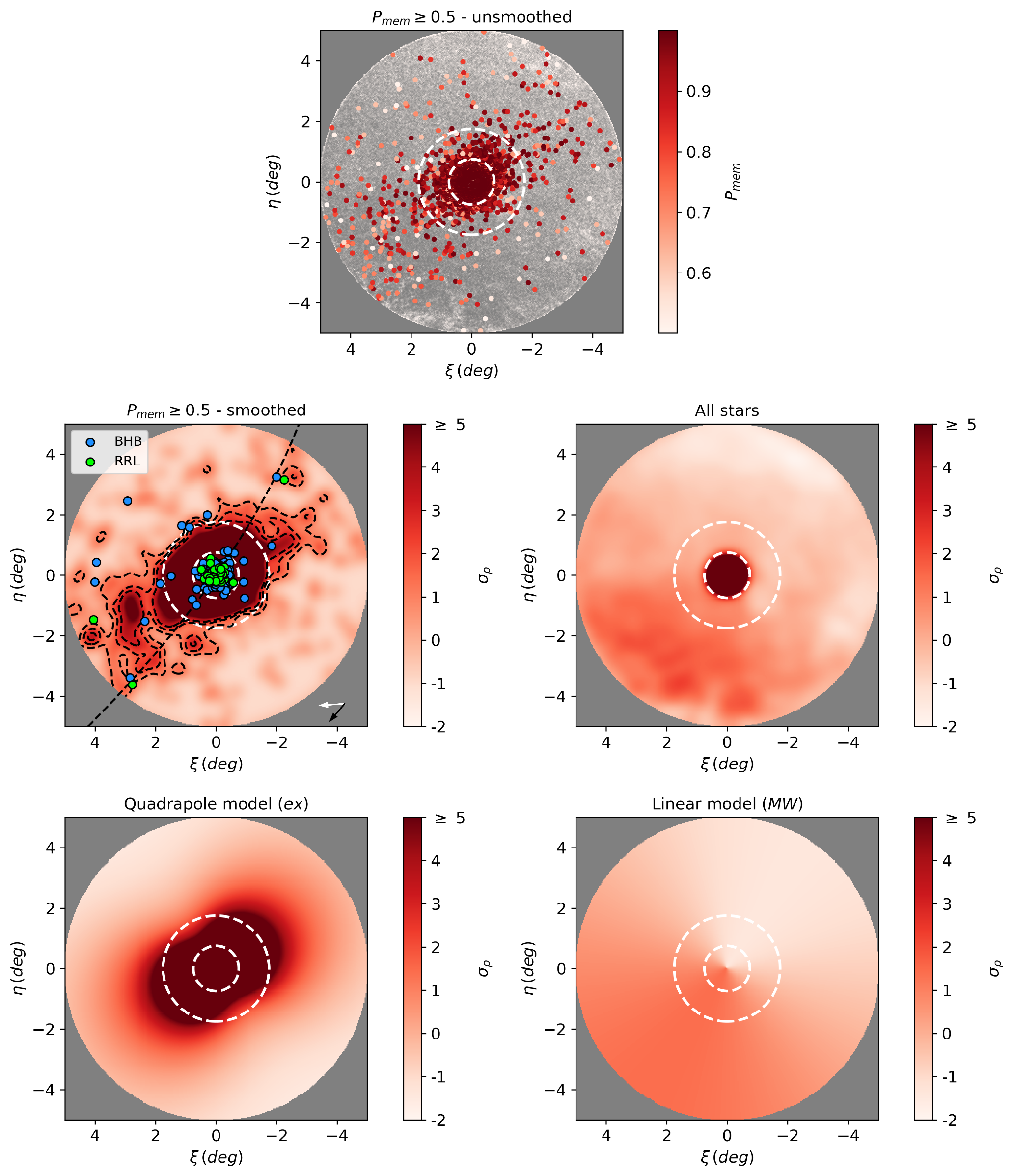}
  \end{center}
\caption{Top row: Unsmoothed spatial distribution of individual stars with $P_{mem} \geq 0.5$, with the colour scale denoting increasing probability. Stars with $P_{mem} < 0.3$ are shown in grey. Middle row: The left figure displays stars with $P_{mem} \geq 0.5$ binned into $3\cross3$ arcmin cells and subsequently smoothed with a Gaussian of width 12 arcmin. Contours here represent $1\sigma$, $2\sigma$ and $3\sigma$ above the mean bin value (outside 1.2 times the Jacobi radius) and the colour scale represents the standard deviation, $\sigma_{\rho}$, about the mean bin density, as described in Section \ref{sect:tidal_extensions}. Candidate $\omega$ Cen blue horizontal branch (BHB, blue) and RR-Lyrae stars (RRL, green) are also shown. The dashed line indicates the orbit of $\omega$ Cen which traverses from North West to South East. The arrows at the bottom of the left panel show the direction of the solar reflex motion-corrected proper motion of $\omega$ Cen (black arrow) and the direction to the Galactic Centre (white arrow). The right figure shows the 500,000 stars from Section \ref{sec:photsel}, prior to the Bayesian analysis, binned and smoothed as in the top right figure. Bottom row: The left figure shows the model of the extended structure (Eq. \ref{eq:qudspat}) with our parameters. The direction of the quadrupole describes the data well.  The right figure shows the fitted linear model (Eq. \ref{eq:linspat}) to the contamination. In all cases, the inner and outer white dashed circles indicate the King tidal radius of $46.4$ arcmin and Jacobi radius of $106$ arcmin respectively.}
\label{fig:2ddens}
\end{figure*}

The morphology of the extended component is visualised in Fig. \ref{fig:2ddens} which shows the surface density distribution of stars with $P_{mem}\geq 0.5$.  Stars are assigned to 3 arcmin $\cross$ 3 arcmin bins on the sky and subsequently smoothed with a Gaussian of width 12 arcmin. We calculated the mean bin density of stars lying outside 1.2 times the Jacobi radius 161.7~pc or $106$ arcmin \citep{2018MNRAS.474.2479B} and the associated spread, hence removing the influence of the cluster itself in our statistics. Each bin is then displayed as the number of standard deviations above the aforementioned mean bin density. For comparison, we also show the distribution of all 500,000 stars, created under the same conditions, that passed our initial selection  criteria prior to the Bayesian analysis. The dominant  gradient across the field seen here has been effectively removed by our technique to reveal the underlying structure.  \ref{fig:2ddens} also shows the model fits to the extended structure (quadrupole) and MW contamination (linear), demonstrating they describe the data well. 

The tidal tails extend across the full extent of the analysed area, corresponding to a physical diameter of $\sim900$~pc.  We have explored greater radial extents surrounding the cluster but find no significant detection of the debris beyond our initial search radius of 5 degrees; however, on such large scales, our assumption of a single value for the cluster proper motion will begin to break down. The tidal tails we have uncovered exhibit the same overall morphology as those seen in other recent studies of $\omega$ Cen \citep{2019NatAs.tmp..258I,2020MNRAS.495.2222S}, as well as those hinted at in earlier work \citep{2014MNRAS.444.3809M}, but have been recovered here through a completely independent method and one which allows membership probabilities to be associated to individual stars.  Fig. \ref{fig:2ddens} also shows the orbital path of $\omega$ Cen calculated with \textsc{Galpy}\footnote{http://github.com/jobovy/galpy}, using the  MWpotential2014 from \cite{2015ApJS..216...29B} and the $\omega$ Cen positional and velocity values from  \citep{2019MNRAS.482.5138B}. There is a reasonably good correspondence between path of the orbit and the direction of the extended structure, which is to be expected as tidal extensions are typically found to closely follow the orbit \citep[e.g.,][]{2007ApJ...659.1212M}.

\subsection{Radial Properties}\label{sect:radial_profile}
\begin{figure}
  \begin{center}  
    \includegraphics[width=\columnwidth]{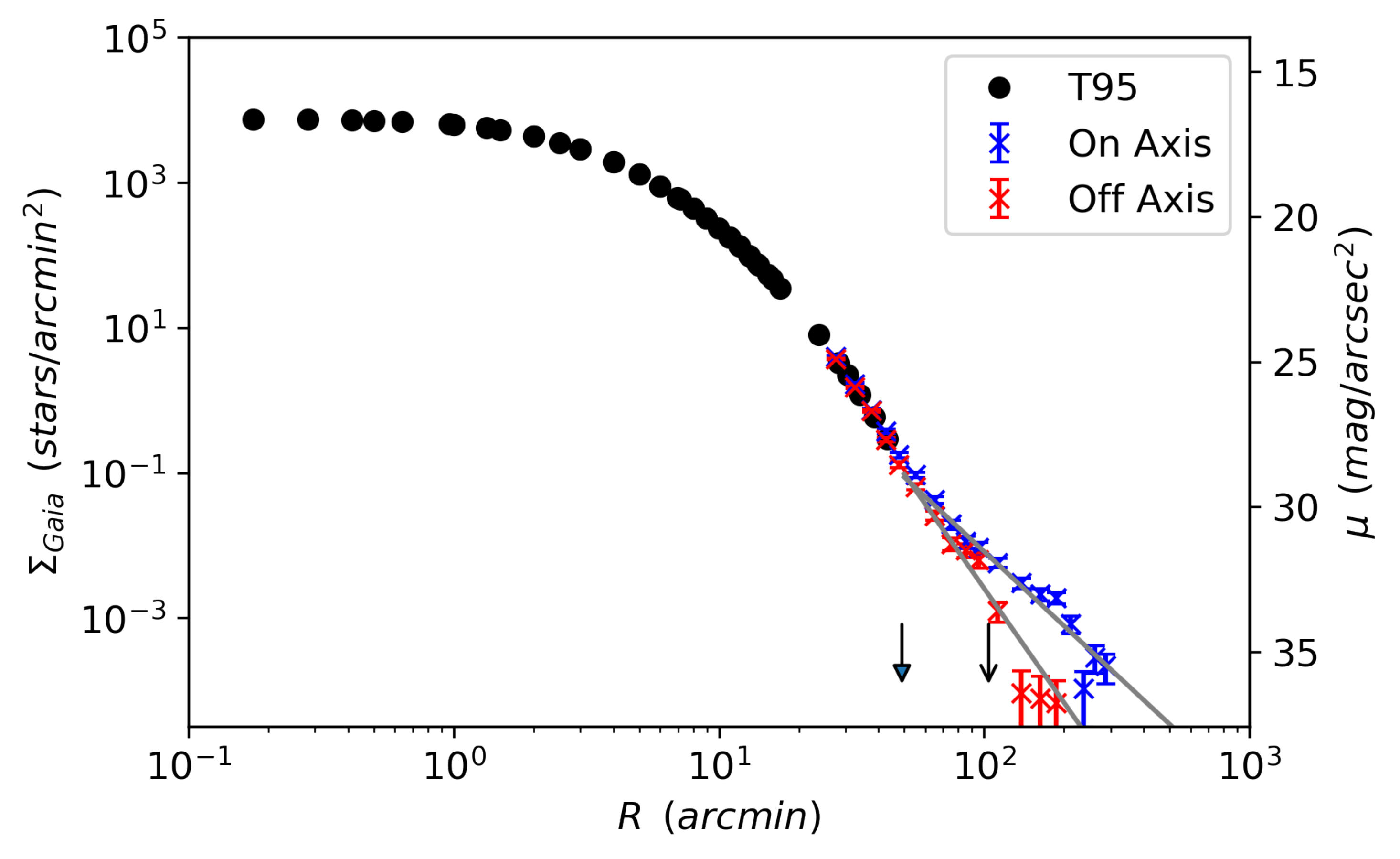}
  \end{center}
\caption{Radial profile of $\omega$ Cen of stars with $P_{mem}\geq 0.5$ within the 2 $\sigma$ detection presented in Fig. \ref{fig:2ddens}. The left y-axis shows density from star counts while the right y-axis demonstrates the corresponding surface brightness. The outer star count profile is shown along both on and off-axis (blue and red points respectively) of the extensions as described in Section \ref{sect:radial_profile}. The King tidal radius and the Jacobi radius are indicated by closed and open arrows respectively. We have supplemented our profile with the inner surface brightness profile from \citet{1995AJ....109..218T} shown by the solid black circles. The solid grey lines show the power-law fits to both the on and off-axis regions, with $\gamma = -3.40$ and $-5.22$ respectively.}
\label{fig:rpprof}
\end{figure}

 Fig. \ref{fig:rpprof} shows the radial fall-off of the extended component derived from counting stars along the direction of the tails (on-axis) and perpendicular to it (off-axis).  For constructing these profiles, we have used only $P_{mem}\geq 0.5$ stars that are within the $2\sigma$ density detection contour of Fig. \ref{fig:2ddens}. The on-axis profiles were calculated using stars that lie within a 45 $\deg$ wedge, centered on $\omega$ Cen, of the position angle defined by $\theta_{ex}$. The off-axis profiles included those stars within a 45 $\deg$ wedge at an angle perpendicular to $\theta_{ex}$. Within these wedges, we counted the number of stars in concentric annuli out to 5 degrees, with the size of the annuli increasing at large radii to combat small number statistics.  Due to issues with incompleteness, we exclude stars in inner 20 arcmin of $\omega$ Cen \citep[see also][]{2019MNRAS.485.4906D} and instead represent the behaviour of the radial profile within this radius using the azimuthally-averaged V-band surface brightness profile of \citet{1995AJ....109..218T}.  The surface brightness profile and star count profiles are stitched together by scaling the datapoints in the 20-40 arcmin range, allowing us to construct a radial profile that spans $\sim$ 20 magnitudes in surface brightness. Poisson uncertainties are shown for the star counts. 
 
 The on and off axis profiles exhibit very similar behaviour to a radial distance of $\sim$ 40 arcmin ($\sim60$ pc),  indicating a predominantly spherical morphology out to the King tidal radius. Beyond this radius, the stellar distribution becomes increasingly elongated along the axis of the tails, as manifest by both a higher surface density of stars at a given circular radius as well as detections further out.  Also shown in Fig. \ref{fig:rpprof} are power-law fits to both the on and off-axis profiles for $P_{mem}\geq 0.5$ stars lying beyond the King tidal radius. For the on-axis profile, we find that the density follows a $\gamma_{\rm on} = -3.40 \pm 0.20$ decline, and a sharper decline for the off-axis profile, $\gamma_{\rm off} = -5.22 \pm 0.26$. These values are consistent with the power-laws seen in clusters with tidal tails originating from mass-loss \citep[e.g.,][]{2014MNRAS.445.2971C,2015ApJ...803...80K}. 

The fraction of the present-day mass in the tidal extensions can be calculated by integrating the radial profile in Fig. \ref{fig:rpprof} using the LIMEPY  code \citep{2015MNRAS.454..576G}. For this calculation, we assume that mass follows light and that the mass function of stars is a constant in these peripheral parts. To facilitate comparison with other work, we use the Wilson truncation radius, found to be 
 70.6 $\pm 1.2$ arcmin (consistent with \citealt{2019MNRAS.485.4906D}), as our fiducial radius.  For reference, this is $\sim1.5$ times the King tidal radius. Taking the average of the on and off-axis spherical integrations, we find that the amount of mass in the radial range from the Wilson truncation radius to the edge of our field, is a mere $0.1$ per cent of the total mass in the system.  Interestingly, this is considerably less than the $\sim 1$ per cent mass fractions contained in the stellar envelopes seen around NGC~1851 and NGC~7089, which have been calculated in an identical manner  \citep{2018MNRAS.473.2881K}.
 It is also far less than the mass fractions of 3--50 per cent seen in the tidal tails of some lower mass 
 MW globular clusters, such as Palomar~5 \citep{2003AJ....126.2385O}, NGC~5466 \citep{2006ApJ...639L..17G} and M92 \citep{2020ApJ...902...89T}.
 However, it is worth noting that our estimate is based on only the stream stars which lie within a five degree radius of the cluster, and hence is likely to be a lower limit on the fractional mass in the extensions. 
 
\begin{figure*}
 \begin{center}  
    \includegraphics[width=0.9\textwidth]{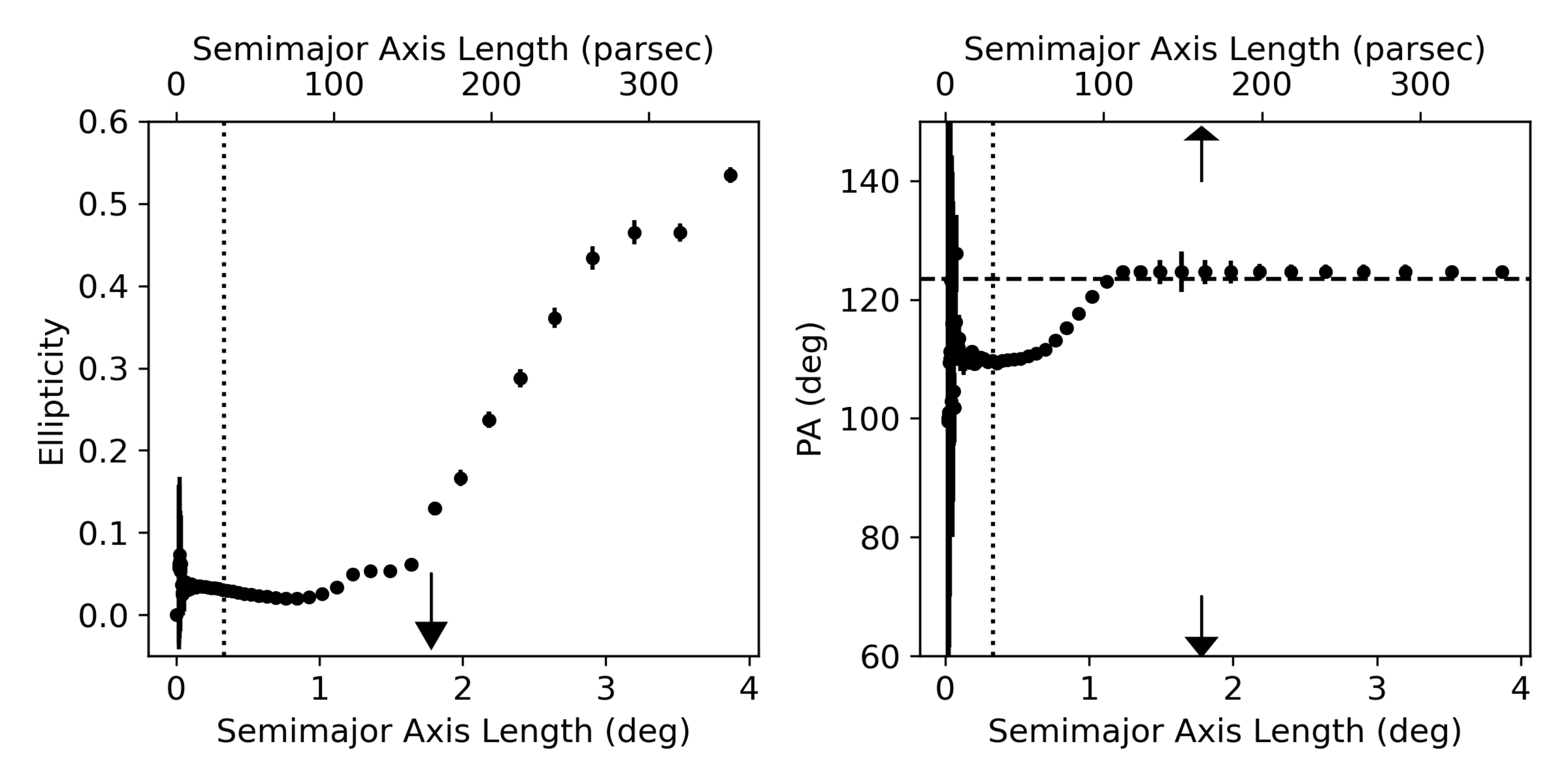}
  \end{center}
\caption{The radial variation of the ellipticity (left) and position angle (right) as determined from ellipse fits to the density distribution of $P_{mem}\geq 0.5$ members. The dotted vertical line in both plots indicates a radius of 20 arcmin where Gaia incompleteness becomes severe and our results are unreliable. The arrows indicate the location of the Jacobi radius. The horizontal dashed line shows $\theta_{ex}$ as determined from our Bayesian analysis. The ellipticity starts to increase sharply at around 1.5 degrees, while the position angle remains constant in this region. }
\label{fig:posa_ell}
\end{figure*}

We also explore how the ellipticity and position angle of $\omega$ Cen vary as a function of radius from the main body into the tails. To do this, we fit elliptical isophotes to the 2D surface density distribution displayed in Fig. \ref{fig:2ddens} using the fitting routine in the python package, \textsc{photutils}. This package performs the fitting based on the iterative method introduced by  \cite{1987MNRAS.226..747J} and in our analysis we choose to hold the centre fixed. The resulting profiles are presented in Fig. \ref{fig:posa_ell} and show that $\omega$ Cen remains roughly spherical to a radius of about 1.5 degrees, which corresponds to just inside the Jacobi radius, before becoming increasingly elliptical at larger distances. The position angle converges on $\theta_{ex}$ at approximately 1.5 degrees as well.

 Prior to our work, the ellipticity of $\omega$ Cen had only been explored out to 30 arcmin, which is well within the King tidal radius.  
 Within this radius, \cite{2020ApJ...891..167C} have shown that the cluster becomes increasingly elliptical as the radius decreases, peaking at a value of $\sim 0.16$ at 8~arcmin. Our measurements at radii $\lesssim 20$~arcmin are unreliable due to incompleteness issues with Gaia but in the range of $20-30$~arcmin we are in excellent agreement with \cite{2020ApJ...891..167C}. Future data releases of Gaia will be better suited to deal with the highly crowded regions of GCs and will allow homogeneous study of the ellipticity and position angle of $\omega$ Cen from its inner regions to furthest extent.
 
\subsection{Other Stellar Population Tracers}
 In our analysis,  we have only considered cluster stars that lie on the upper  main sequence, main sequence turn-off and RGB regions of the CMD. However, Fig. \ref{fig:CMD} demonstrates that $\omega$ Cen possesses a prominent horizontal branch as well, composed of BHB and RRL stars. To examine whether the distribution of these populations is also consistent with the tidal extensions, we revisit our final sample of 
 $5\times10^{5}$ stars described in Section \ref{sec:photsel} and select those objects with $14 \lesssim G_0\lesssim 16$ and $-0.3 \lesssim (G_{bp}-G_{rp})_0\lesssim 0.4$. We also require that these stars have proper motions that lie within 3 sigma of the cluster value listed in Table \ref{tab:SDSS}.  We verified that those objects remaining have parallaxes consistent with that of $\omega$ Cen. Fig. \ref{fig:2ddens} shows the locations of these candidate BHB stars superposed on the surface density distribution of main sequence and RGB stars. The majority of the BHB candidates are confined to the main body of the cluster, which is to be expected given that it is the lower mass (i.e., main sequence) stars that are preferentially populating the tidal extensions \citep{2018MNRAS.474.2479B}.  However, those BHB candidates that do lie  outside the cluster region follow the tidal extensions rather closely.  
 
 To explore the RRL star distribution, we relied on the tables \textit{vari\_rrlyrae} and \textit{vari\_classifier\_result} from the Gaia DR2 analysis of \citet{2018A&A...618A..30H}.  We selected sources that lie within a five degree radius of $\omega$ Cen and, as for the BHBs,  that have proper motions that lie within 3 sigma of the cluster value. As shown in Fig. \ref{fig:2ddens}, the RRL candidates are also highly clustered in the central regions but the three objects which lie beyond the Jacobi radius show a good alignment with the tidal features.  For these three objects, we calculate an time-averaged dereddened magnitude of $G_0=14.2 \pm 0.02$.  Assuming the absolute magnitude of RRL (type AB) to be  $M_G=0.64 \pm 0.25$  \citep{2019MNRAS.482.3868I}, we calculate the distance to these stars to be 
 $5.2 \pm 0.6$ kpc, which is entirely consistent with the measured distance of  $\omega$ Cen \citep{1996yCat.7195....0H,2021ApJ...908L...5S}.  The small number of candidate RRL stars that we have found coincident with the $\omega$ Cen tidal debris is in agreement with the earlier work of \citet{2015A&A...574A..15F}, who failed to find any strong RRL candidates in a search area of 50 sq. degrees around the cluster. Of the small number of potential candidates they identified beyond the tidal radius, we have used EDR3 to confirm that none of these stars have proper motions consistent with $\omega$ Cen membership.

\subsection{Comparison to Radial Velocity Studies}

Our analysis is based on a mixture model approach for spatial positions and proper motions, applied to a sample of stars that have been pre-selected on photometric properties. In this section, we examine what radial velocity (RV) data exist for our sample and whether these measurements are consistent with our membership assignments. 

The RV of $\omega$ Cen is well-determined and large  
 \citep[234 km s$^{-1}$][]{2019MNRAS.482.5138B}, meaning that there is excellent contrast with the surrounding field population. Gaia DR2 (and EDR3) includes mean RVs for 7.2 million stars brighter than $G\approx 13$ and with $T_{\rm eff}$ in the range $\sim 3550-6900$~K \citep{2016A&A...595A...1G,2020arXiv201201533G}.  We explored which of our sample of $5\times10^{5}$ stars in the vicinity of $\omega$ Cen have Gaia RVs.  Of the 48 stars that do, we find that 13 of these stars have $P_{mem}\approx 1$ while the remaining 35 have $P_{mem}\approx 0$. Reassuringly, the $P_{mem}\approx 1$ stars all have RVs which lie within 10~km s$^{-1}$ of the mean $\omega$ Cen velocity whereas the $P_{mem}\approx 0$ all have RVs $\lesssim 170$~km s$^{-1}$, further confirming they are non-members.  All the stars with Gaia RVs lie within 30 arcmin of the cluster centre. 
 
 We also searched other public spectroscopic survey datasets for RVs for our sample and found that 295 of our stars have RVs in APOGEE DR16 \citep{2020AJ....160..120J}.  Of these, we find 291 have $P_{mem}\approx 1$ and RVs consistent with cluster membership but that only two of these lie beyond 30 arcmin radius.  These numbers are unsurprising given that the APOGEE-2 observations come from a targetted program on $\omega$ Cen but it is still encouraging that so many are returned as high probability members by our method. 
 
There have been very few dedicated spectroscopic searches for stars in the far outer regions of $\omega$ Cen stars thus far.   \citet{2008AJ....136..506D} and \citet{2012ApJ...751....6D} (hereafter DC12) used the AAT 2dF to search for candidate stars on the lower red giant branch over the radial range 20-60 arcmin, i.e. from roughly half to just beyond the King tidal radius. We cross-matched the probable members and probable non-members from DC12 with our catalogue. We find that our sample contains 102 out of the 160 DC12 probable members, of which we confirm 99 as high probability members ($P_{mem}>0.5$). Those not present in our sample, as well as those not recovered as high probability members,  are all located within the inner 30 arcmin of the cluster,  where crowding affects the astrometry and photometry and where the multiple population signature causes some stars to be missed by our photometric selection. Interestingly, our technique also identifies 11 (~30 per cent) of DC12's probable non-members as stars with $P_{mem}>0.5$. These turn out to be stars that have RVs consistent with that of $\omega$ Cen but that DC12 suspected were contaminants on the basis of their line-strengths. Our analysis indicates that these 11 stars are indeed members of the cluster. All of the cross-matched member stars lie within the tidal radius.

Finally, \citet{2009MNRAS.396.2183S} conducted a search using VLT/FLAMES for $\omega$ Cen stars at large radii.  Cross-matching our catalogue with their RV members, we find 90 stars in common, all of which have $P_{mem}\approx 1$ and lie within 35 arcmin of the cluster centre. 

The fact that our algorithm can independently recover known RV members with (very) high probability is reassuring but unfortunately the comparisons we have been able to make are quite limited due to the fact that there are so few known RV members beyond the tidal radius. Indeed, the majority of targets we have identified as members of the tidal features are fainter than the typical depth of existing spectroscopic surveys and the heavy contamination along this sightline means that previous targetted programmes have been inefficient in identifying RV members at large radii. Our sample of $\approx4000$ high probability candidates outside 30 arcmin provides an excellent sample for future spectroscopic follow-up, which will enable confirmation of membership as well as measurements of RVs and chemistry. 

 \section{Conclusions}
 We have searched for extended tidal structure surrounding $\omega$ Cen using a mixture model technique that is solved within a Bayesian framework. Our approach is unique compared to previous work of this type in that we separately model the cluster, the extended component and the field contamination. Furthermore, our modelling has the flexibility to detect both symmetric tidal tails as well as extended spherical envelopes at large radii.  Our analysis,  which utilises photometry and astrometry from Gaia EDR3, yields membership probabilities for stars out to 5 degrees from the center of $\omega$ Cen, corresponding to a physical distance of 450~pc.  Examining the distribution of high probability ($P_{mem}\geq 0.5$) members across our field-of-view reveals spectacular tidal extensions, independently confirming results seen in earlier work \citep{2019NatAs.tmp..258I, 2020MNRAS.495.2222S} that adopted different methods. We have used this sample to analyse the structure of the peripheral regions of $\omega$ Cen, characterising the radial fall-off along the major and minor axes, as well as the radial variation of the ellipticity and the position angle.  We have shown that both RR Lyrae stars and BHB stars consistent with belonging to $\omega$ Cen are found to exist along the extensions and that the tails constitute only a small fraction ($\approx0.1\%$) of the overall cluster stellar mass.    
Our high probability members provide prime targets for future spectroscopic studies of $\omega$ Cen out to unprecedented radii.  Indeed, almost all RV and chemical abundance analyses to date have focused on stars in the inner regions of the cluster, with only a small handful of RVs measured out to the King tidal radius \citep{2008AJ....136..506D,2009MNRAS.396.2183S}. Establishing the kinematics and especially the chemistry of the outlying populations we have uncovered will be crucial for firming up the link between $\omega$ Cen and the \textit{Fimbulthul} stream, as well as for providing a direct measure of the chemical composition of the cluster's stars that escape into the MW halo. Such data are also likely to shed further light on the specific accretion event that brought $\omega$ Cen into the MW.  

The proximity of $\omega$ Cen to the Galactic plane makes this object 
arguably one the most challenging GCs for studies of extra-tidal populations.  This paper has demonstrated the efficacy of our probabilistic technique to recover faint structure in its outer parts. 
 In future contributions, we will present results from a much larger sample of GCs in which we will address the statistical properties of GC extensions.  Follow-up spectroscopy of these regions will be required to fully characterise and interpret the significance of extra-tidal features; fortunately, this is a task that is very well suited to forthcoming wide-field high-multiplex facilities such as WEAVE \citep{2012SPIE.8446E..0PD} and 4-metre Multi-Object Spectrograph \citep[4MOST;][]{2019Msngr.175....3D}.

\section*{Acknowledgements}
This work makes use of the following software packages: \textsc{astropy} \citep{2013A&A...558A..33A,2018AJ....156..123A}, \textsc{astroquery} \citep{2019AJ....157...98G}, \textsc{corner} \citep{corner}, \textsc{dustmaps} \citep{2018JOSS....3..695M}, \textsc{Gala} \citep{gala,adrian_price_whelan_2020_4159870}, \textsc{matplotlib} \citep{Hunter:2007}, \textsc{numpy} \citep{2011CSE....13b..22V}, \textsc{PhotUtils} \citep{2020zndo...4049061B}, \textsc{PyMultinest} \citep{2014A&A...564A.125B}, \textsc{scipy} \citep{2020SciPy-NMeth}.

PBK is grateful for support from a Commonwealth Rutherford Fellowship from the Commonwealth Scholarship Commission in the UK as well as from UKRI (MR/S018859/1). We gratefully acknowledge useful discussions with Geraint Lewis, Joe Zuntz and especially Anna Lisa Varri while developing this work. We also thank the referee who provided very useful comments. This work has made use of the resources provided by the Edinburgh Compute and Data Facility (ECDF) (http://www.ecdf.ed.ac.uk/).\\
This work has made use of data from the European Space Agency (ESA) mission
{\it Gaia} (\url{https://www.cosmos.esa.int/gaia}), processed by the {\it Gaia}
Data Processing and Analysis Consortium (DPAC,
\url{https://www.cosmos.esa.int/web/gaia/dpac/consortium}). Funding for the DPAC
has been provided by national institutions, in particular the institutions
participating in the {\it Gaia} Multilateral Agreement.

\section*{Data Availability}
The list of highly-probable members of $\omega$ Cen which underpins this article and corresponding figures will be shared on reasonable request to the corresponding author.

\bibliographystyle{mnras.bst}

\bibliography{Omega_tails}
\bsp	
\label{lastpage}
\end{document}